\documentclass[aps,prl,reprint,superscriptaddress]{revtex4-2}
\usepackage[T1]{fontenc} 

\usepackage{amsmath}
\usepackage{graphicx} 

\begin{document}
	
	
	\title{ Testing the Reptation Picture: Topological Constraint from Monomer Dynamics }
	
	
	\author{Xiaofei Tian}
	\affiliation{School of Chemical Engineering and Light Industry, Guangdong University of Technology, Guangzhou 510006, P. R. China}
	\author{Qinhang Liu}
	\affiliation{School of Chemical Engineering and Light Industry, Guangdong University of Technology, Guangzhou 510006, P. R. China}
	\author{Zhi-Chao Yan}
	\affiliation{School of Chemical Engineering and Light Industry, Guangdong University of Technology, Guangzhou 510006, P. R. China}
	\author{Liang Gao}
	\affiliation{School of Chemical Engineering and Light Industry, Guangdong University of Technology, Guangzhou 510006, P. R. China}
	\author{Tongfei Shi}
	\affiliation{School of Chemical Engineering and Light Industry, Guangdong University of Technology, Guangzhou 510006, P. R. China}
	\author{Jizhong Chen}
	\email{jzchen@gdut.edu.cn}
	\affiliation{School of Chemical Engineering and Light Industry, Guangdong University of Technology, Guangzhou 510006, P. R. China}

	
	\date{\today}
	
	\begin{abstract}

	The reptation model postulates that entangled polymers slide within a fractal tube. Here we employ a model-independent relation between the zero-displacement probability and the mean-square displacement that applies to time-dependent fractal structures, enabling direct measurement of the fractal dimension $d_\mathrm{f}$ of the geometry experienced by monomer motion. For two-dimensional obstacle arrays and in the slip-link model, $d_\mathrm{f}$ agrees with the reptation prediction $d_\mathrm{f}=1/\nu$ (where $\nu$ is the Flory exponent). In polymer melts, however, we find $d_\mathrm{f} \approx 2.6$ --- a value close to the fractal dimension of percolation clusters, not the reptation value $d_\mathrm{f}=2$. This contrasts sharply with the reptation picture, in which a Rouse chain slides in a fractal structure with $d_\mathrm{f}=2$, spectral dimension $d_\mathrm{s}=1$, and walk dimension $d_\mathrm{w}=4$; our results point instead to a percolation-like scenario, characterized by $d_\mathrm{f}\approx 2.6$, $d_\mathrm{s}\approx 1.3$, and $d_\mathrm{w}\approx 4$ --- revealing a dynamically emergent, finite-size fractal geometry distinct from the static tube.

	\end{abstract}
	
	
	\maketitle
	
	
	\sloppy 

	Pioneered by de Gennes~\cite{de1971reptation, de1979scaling} and extended by Doi and Edwards~\cite{doi1978dynamics1, doi1978dynamics2, doi1978dynamics3, doi1979dynamics4, doi1988theory}, the reptation model has long served as the cornerstone of entangled polymer dynamics. It envisions a chain sliding along the contour of a fractal tube formed by the surrounding chains (Fig. 1(a)). This conceptual picture has been remarkably successful: it not only captures essential dynamical features such as multistage monomer subdiffusion~\cite{wischnewski2003direct} and the chain-length dependence of terminal relaxation~\cite{kremer1990dynamics}, but also profoundly shapes our understanding of the linear~\cite{likhtman2002quantitative, milner1998reptation} and nonlinear~\cite{mhetar1999nonlinear, morse1999viscoelasticity} rheology of entangled polymers. Building on this picture, various simulation-based analysis techniques have been developed to identify the constraints imposed by neighboring chains~\cite{everaers2004rheology, kroger2005shortest, shanbhag2007primitive, kroger2023z1plus, tzoumanekas2006topological, anogiannakis2012microscopic}, thereby visualizing the static entanglement network. However, in melts, a chain moves in concert with its surrounding chains, and the central question --- whether the constraints thus imposed can indeed be equated to static obstacles, i.e., whether the confining geometry experienced by a monomer’s motion is truly a tube --- has never been directly tested. Even with important refinements such as constraint release (CR)~\cite{rubinstein1988self, viovy1991constraint}, this issue persists, as such mechanisms do not alter the underlying static-obstacle picture.

	The primary challenge in probing the confining geometry experienced by a monomer’s motion lies in its inherently temporal nature. The reptation model provides a comprehensive description: a monomer begins to sense topological constraints at the entanglement time $\tau_\mathrm{e}$, subsequently moves within a confining tube of fractal dimension $d_\mathrm{f}$, and eventually escapes from its original tube at the terminal relaxation time $\tau_\mathrm{d}$. Consequently, methods developed for static fractal structures are not directly applicable to identifying such a time‑dependent geometry. One common indirect approach is to estimate $d_\mathrm{f}$ via the Alexander–Orbach (AO) relation $d_\mathrm{f}/d_\mathrm{w}=d_\mathrm{s}/2$~\cite{alexander1982density}, which connects the walk dimension $d_\mathrm{w}$ (obtained from the mean‑square displacement (MSD) scaling $\langle\Delta r^{2}(t)\rangle\sim t^{2/d_\mathrm{w}}$) to the spectral dimension $d_\mathrm{s}$ (derived from the zero‑displacement probability scaling $P(0,t)\sim t^{-d_\mathrm{s}/2}$). Reliable application of this relation requires clear power‑law regimes in both quantities. In entangled polymers, however, this approach is severely hampered by the intrinsic multi‑scale nature of the dynamics. Determining stable power‑law exponents $d_\mathrm{w}$ and $d_\mathrm{s}$ becomes inherently ambiguous due to pervasive dynamical crossovers. It is well established that the MSD of a tagged monomer exhibits complex scaling behavior~\cite{azuma1999diffusion}:
	\begin{equation}
		\langle\Delta r^{2}(t)\rangle \sim \left\{\begin{matrix}
			t^{2\nu/(1+2\nu)},\quad\quad & \tau_{0} < t < \tau_\mathrm{e},\\[0.8ex]
			t^{\nu/2}, \quad\quad\qquad & \tau_\mathrm{e} < t < \tau_\mathrm{R},\\[0.8ex]
			t^{\nu},\quad\quad\quad\qquad & \tau_\mathrm{R} < t < \tau_\mathrm{d},\\[0.8ex]
			t,\quad\quad\quad\qquad & t > \tau_\mathrm{d}.\quad\quad
		\end{matrix}\right.
		\label{001}
	\end{equation}
	where $\tau_0$ is the monomer relaxation time, $\tau_\mathrm{R}$ the Rouse time, and $\nu$ the Flory exponent. Importantly, $d_\mathrm{f}$ itself is expected to vary across $\tau_\mathrm{e}$ and $\tau_\mathrm{d}$, rather than to follow the evolving MSD exponent. Moreover, the finite extent of this temporal fractal structure leads to sharp changes in $P(0,t)$: once the monomer exits the low‑dimensional space, it rarely returns to previously visited sites. Consequently, power-law behavior in $\langle\Delta r^{2}(t)\rangle$ and $P(0,t)$ exists only within limited time windows, rather than across the entire entanglement regime, making the AO relation difficult to apply reliably. This limitation motivates the development of an alternative approach that does not presuppose steady‑state fractal scaling.
		\begin{figure}[t]
		\centering 
		\includegraphics[width=\linewidth]{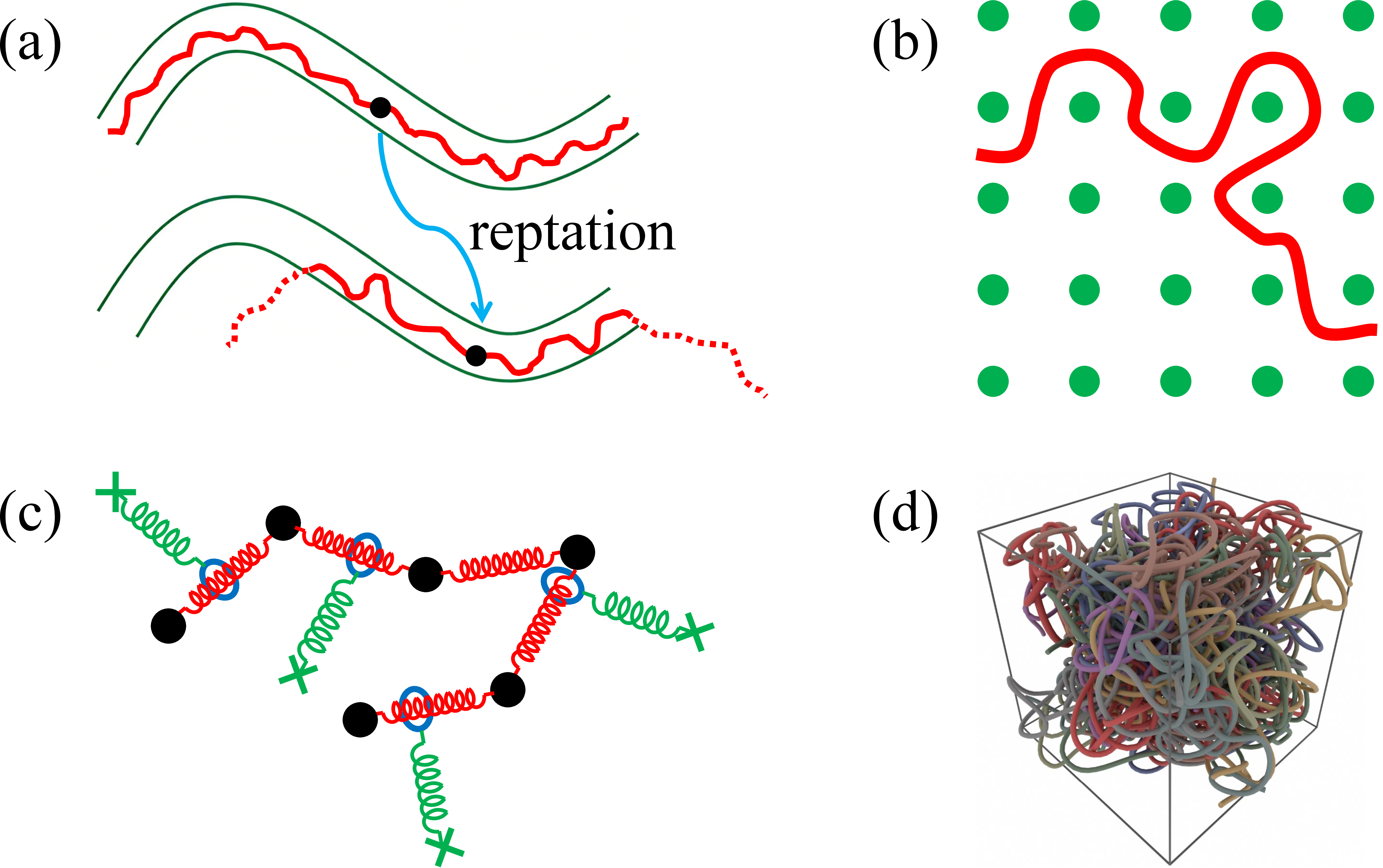} 
		\caption{\label{fig1} Schematic: (a) the reptation picture; (b) a polymer in a $2$D quenched obstacle array; (c) the Likhtman slip-link model; (d) an entangled polymer melt.}
	\end{figure}	

	In this Letter, we measure the fractal dimension $d_\mathrm{f}$ of the geometry experienced by monomer motion, providing a direct characterization of the time-dependent fractal structures imposed by topological constraints. We first derive a relation between the zero-displacement probability and the MSD that yields $d_\mathrm{f}$ --- a robust, model-independent probe of the temporal confining geometry. We then apply it to molecular dynamics simulations of three representative entangled systems: a chain in a quenched obstacle array~\cite{nixon1999relaxation} (Fig.~\ref{fig1}(b)), which directly embodies de Gennes' static-tube premise; the Likhtman slip-link model~\cite{likhtman2005single} (Fig.~\ref{fig1}(c)), a tube-based model that can incorporate constraint release; and entangled polymer melts (Fig.~\ref{fig1}(d)). This systematic strategy provides a direct test of the fractal tube premise. Combined with independent measurements of the spectral dimension $d_\mathrm{s}$ and the walk dimension $d_\mathrm{w}$, we fully characterize the time-dependent fractal structures governing monomer motion.

	We present a direct method to determine the fractal dimension $d_\mathrm{f}$ from its geometric definition. By treating the continuous space explored by a tracer as a set of discrete sites, the fractal dimension is defined through the scaling of the number $M$ of accessible sites with radius $r$: $M \sim r^{d_\mathrm{f}}$~\cite{meroz2013test, meroz2015toolbox}. Substituting $r$ with the root-mean-square displacement $\langle \Delta r^{2}(t)\rangle^{1/2}$, which characterizes the explored region up to time $t$, gives $M \sim \langle \Delta r^{2}(t)\rangle^{d_\mathrm{f}/2}$. Let $S_t$ denote the total number of distinct sites visited by time $t$. For a recurrent random walk, all sites in the explored region are eventually visited, so $S_t\approx M$~\cite{meroz2013test, dasgupta1994distinct}, giving 
		\begin{equation}
		S_t  \sim \langle \Delta r^{2}(t)\rangle^{d_\mathrm{f}/2}
		\label{002}
	\end{equation}
	Since $P(0,t)$ is inversely proportional to the number of distinct sites visited,
	\begin{equation}
		P(0,t) \sim S_t^{-1}
		\label{003}
	\end{equation}
	we obtain the key relation~\cite{alexander1982density}
	\begin{equation}
		P(0,t) \sim \langle \Delta r^{2}(t)\rangle^{-d_\mathrm{f}/2}
		\label{004}
	\end{equation}
	which provides a direct route to extract $d_\mathrm{f}$ from simultaneous measurements of $P(0,t)$ and $\langle \Delta r^{2}(t)\rangle$ without requiring the identification of distinct temporal scaling regimes --- thereby overcoming the principal limitation of the AO relation. When the displacement distribution is Gaussian and $d_\mathrm{f}=d$ (with $d$ being the Euclidean dimension), $P(0,t)$ reduces to the familiar form~\cite{meroz2015toolbox} 
	\begin{equation}
		P(0,t) = \left[\frac{2\pi \langle \Delta r^{2}(t)\rangle}{d} \right]^{-d/2}
		\label{005}
	\end{equation}
	 Notably, if the scaling behaviors $P(0,t) \sim t^{-d_{s}/2}$ and $\langle \Delta r^{2}(t)\rangle \sim t^{2/d_\mathrm{w}}$ hold in a given dynamical regime, Eq.~\eqref{004} reduces exactly to the AO relation $d_\mathrm{f}/d_\mathrm{w}=d_\mathrm{s}/2$. Thus, our relation provides a general criterion for assessing the validity of AO scaling in specific dynamical regimes.
	
	It is instructive to examine the fractal dimension $d_\mathrm{f}$ of the constraint structure predicted by the tube model. In this framework, an entangled chain moves along its contour --- the primitive path --- with curvilinear displacement $x(t)$. At times shorter than the entanglement time ($t<\tau_\mathrm{e}$), the root-mean-square displacement $\langle \Delta r^{2}(t)\rangle^{1/2}$ is smaller than the tube diameter $a^{\ast}$; the accessible sites then fill the entire volume within that radius, so that $S_t \sim \langle \Delta r^{2}(t)\rangle^{d/2}$ and hence $d_\mathrm{f}=d$. During the intermediate regime ($\tau_\mathrm{e}<t<\tau_\mathrm{d}$), the chain explores the tube contour. The spatial displacement of a tagged monomer scales as $\langle \Delta r^{2}(t)\rangle^{1/2}\approx a^{\ast}(x(t)/a^{\ast})^{\nu}$, reflecting the Flory statistics of the primitive path, while the number of accessible sites grows as $S_t \sim (a^{\ast})^{d-1}x(t)\sim \langle \Delta r^{2}(t)\rangle^{1/2\nu}$, yielding $d_\mathrm{f}=1/\nu$. For times longer than the disengagement time ($t>\tau_\mathrm{d}$), the chain escapes the tube and again explores the full spatial volume, restoring $d_\mathrm{f}=d$. This theoretical analysis provides clear, quantitative signatures that can be directly compared with simulation results.
	
	To systematically test the reptation picture, we perform molecular dynamics simulations of three representative models of entangled dynamics, illustrated schematically in Fig.~\ref{fig1}(b)–(d). To eliminate end effects~\cite{wang2012segmental,abadi2018entangled}, we focus on the central monomer of each chain in our analysis. For the $2$D obstructed system and the polymer melts, non-bonded interactions are modeled by the purely repulsive Weeks–Chandler–Andersen (WCA) potential~\cite{weeks1971role}, implemented as a truncated and shifted Lennard-Jones potential with cutoff $r_\mathrm{c}=2^{1/6}\sigma$, where $\sigma$ sets the unit of length. Consecutive monomers along a chain are connected by finitely extensible nonlinear elastic bonds~\cite{kremer1990dynamics}. In the $2$D obstructed system, obstacles of size $\sigma$ are placed on a quenched square lattice with spacing $a=6\sigma$; monomer–obstacle interactions are also described by the repulsive WCA potential~\cite{nixon1999relaxation}. For the Likhtman slip‑link model, we simulate a Rouse chain of phantom beads connected by harmonic springs. Before applying our geometric probe [Eq.~\eqref{004}] to entangled polymers, we validate it on a minimal reference system: a tracer particle confined in a $2$D straight channel. The extracted fractal dimension $d_\mathrm{f}$ exhibits the expected crossover from the spatial dimension ($d_\mathrm{f}=2$) to the channel dimension ($d_\mathrm{f}=1$) once the MSD reaches the channel width, and finnally returns to the spatial dimension $d_\mathrm{f}=2$ after escaping from the channel. This benchmark clearly illustrates the evolution of $P(0,t)$ as a function of $\langle \Delta r^{2}(t)\rangle$ and $t$ as well as the evolution of  $\langle \Delta r^{2}(t)\rangle$ as a function of $t$, throughout the complete process of entering, moving within, and escaping from a finite-size structure. The results aid in analyzing entangled systems. More details are provided in the Supplemental Material (SM)~\cite{SM}. 
	\begin{figure}[t]
		\centering 
		\includegraphics[width=\linewidth]{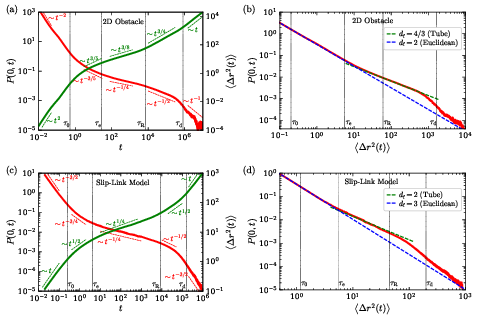} 
		\caption{\label{fig2} (a),(c) $P(0,t)$ (green) and $\langle \Delta r^{2}(t)\rangle$ (red) as functions of $t$; dashed lines indicate the power laws predicted by reptation [Eq.~\eqref{001}] and the AO relation, respectively. (b), (d) $P(0,t)$ versus $\langle \Delta r^{2}(t)\rangle$; dashed lines correspond to the Gaussian prediction [Eq.~\eqref{005}] and the reptation expectation $d_\mathrm{f}=1/\nu$, respectively. The chain lengths are $N=200$ for the $2$D quenched obstacle system and $150$ for Likhtman slip-link model.}
	\end{figure}
	
	We first examine the reptation picture using a real chain diffusing in a $2$D quenched obstacle array. We show $P(0,t)$ and $\langle \Delta r^{2}(t)\rangle$ as functions of $t$ in Fig.~\ref{fig2}(a), and  $P(0,t)$ versus $\langle \Delta r^{2}(t)\rangle$ in Fig.~\ref{fig2}(b). In this system, inertial effects are included in the equations of motion, giving ballistic scaling $\langle \Delta r^{2}(t)\rangle \sim t^{2}$ for $t < \tau_{0}$. Figure~\ref{fig2}(a) shows good agreement between the simulation data and the theoretical scaling obtained by substituting the Flory exponent $\nu = 3/4$ into Eq.~\eqref{001}. The distinct scaling regimes allow us to estimate the characteristic times $\tau_\mathrm{e}$, $\tau_\mathrm{R}$, and $\tau_\mathrm{d}$ from the intersections of the corresponding slopes. At short times ($t < \tau_\mathrm{e}$), the probability density $P(0,t)$ can be described by Eq.~\eqref{005}, corresponding to $d_\mathrm{f}=d=2$ and Gaussian displacement statistics, see Fig.~\ref{fig2}(b). Notably, although both $P(0,t)$ and $\langle \Delta r^{2}(t)\rangle$ exhibit an apparent crossover around $\tau_{0}$ when plotted against $t$, Eq.~\eqref{004} still accurately yields $d_\mathrm{f}$, demonstrating its robustness. The first transition of the fractal dimension from $d_\mathrm{f}=2$ to $d_\mathrm{f}=1/\nu = 4/3$ occurs around $\langle \Delta r^{2}(\tau_\mathrm{e})\rangle$, where the monomer first senses the confining tube, consistent with the reptation picture. This scaling $P(0,t) \sim \langle \Delta r^{2}(t)\rangle^{-2/3}$ persists for $t$ appreciably larger than $\tau_\mathrm{R}$, even after the time dependences of $P(0,t)$ and $\langle \Delta r^{2}(t)\rangle$ have changed, indicating that the monomer remains within the tube. Upon tube escape, $d_\mathrm{f}$ returns from $4/3$ to $2$, accompanied by an apparent crossover regime in $P(0,t)$ versus $\langle \Delta r^{2}(t)\rangle$ around $\tau_\mathrm{d}$. This behavior arises from finite-size effects of the fractal tube: once the monomer exits, it rarely revisits previously occupied sites, causing a sharp drop in the time dependence of $P(0,t)$. Thus, the power-law dependence of $P(0,t)$ on $\langle \Delta r^{2}(t)\rangle$ provides a clear and continuous dynamical signature, capturing the crossover from $2$D diffusion at sub-tube scales, to $1$D curvilinear motion along the tube, and finally to unconstrained diffusion after escape. This progression offers direct geometric confirmation of the structural picture assumed in reptation.
	
	As a concrete computational test of the tube picture, we examine the Likhtman slip-link model with CR that incorporates CR. Using overdamped Langevin dynamics, the simulations reproduce the expected diffusive scaling $\langle \Delta r^{2}(t)\rangle \sim t$ for $t < \tau_{0}$. We first verify that the MSD exhibits the four scaling regimes predicted by Eq.~\eqref{001} for an ideal chain with Flory exponent $\nu=1/2$ (see Fig.~\ref{fig2}(c)). As shown in Fig.~\ref{fig2}(d), the displacement distribution is Gaussian in both the short-time ($t <\tau_\mathrm{e}$) and long‑time ($t>\tau_\mathrm{d}$) limits, following Eq.~\eqref{005}, which corresponds to a Euclidean geometry with $d_\mathrm{f}=d=3$. In the intermediate regime $\tau_\mathrm{e}<t<\tau_\mathrm{d}$, however, we observe $P(0,t) \sim \langle \Delta r^{2}(t)\rangle^{-1}$, yielding $d_\mathrm{f}=1/\nu=2$ --- the precise geometric signature of a fractal tube. This result confirms that the slip-link model captures the central premise of reptation: the chain diffuses along a fractal path of dimension $d_\mathrm{f}=2$. Notably, the inclusion of CR affects the terminal relaxation time $\tau_\mathrm{d}$ but does not alter this geometric picture, a conclusion further supported by simulations of the slip-link model without CR~\cite{SM}.
	
	The clear power-law regimes in the time dependences of $P(0,t)$ and $\langle \Delta r^{2}(t)\rangle$ for $\tau_\mathrm{e}<t<\tau_\mathrm{d}$ allow us to further test the validity of the AO relation. In the reptation picture, the entire chain moves as a Brownian particle within the tube during $\tau_\mathrm{R}<t<\tau_\mathrm{d}$; here, the apparent walk dimension $d_\mathrm{w}^\mathrm{app}$ extracted from the time dependence of $\langle \Delta r^{2}(t)\rangle$ equals the intrinsic walk dimension $d_\mathrm{w}$ characterizing diffusion in the fractal structure, yielding $d_\mathrm{w}^\mathrm{app}=d_\mathrm{w}=2/\nu$. In contrast, Rouse dynamics in the regime $\tau_\mathrm{e}<t<\tau_\mathrm{R}$ gives $d_\mathrm{w}^\mathrm{app}=2d_\mathrm{w}=4/\nu$, arising from the coupling of subdiffusive chain motion with the fractal environment. Consequently, if the AO relation holds, the apparent spectral dimension extracted from the time dependence of $P(0,t)$ is expected to be $d_\mathrm{s}^\mathrm{app}=d_\mathrm{s}/2=1/2$ for $\tau_\mathrm{e}<t<\tau_\mathrm{R}$ and $d_\mathrm{s}^\mathrm{app}=d_\mathrm{s}=1$ for $\tau_\mathrm{R}<t<\tau_\mathrm{d}$. For the $2$D quenched obstacle array, clear power laws in both quantities are observed for $\tau_\mathrm{e}<t<\tau_\mathrm{d}$ (see Fig.~\ref{fig2}(a)), yielding $d_\mathrm{s}=1$ and $d_\mathrm{w}=8/3$ as expected. For the Likhtman slip-link model with CR, by contrast, the power-law regimes are evident for $\tau_\mathrm{e}<t<\tau_\mathrm{R}$ (see Fig.~\ref{fig2}(c)), again giving $d_\mathrm{s}=1$. Thus, after accounting for the crossover regimes around $\tau_\mathrm{e}$, $\tau_\mathrm{R}$, and $\tau_\mathrm{d}$ in the time dependences of $P(0,t)$ and $\langle \Delta r^{2}(t)\rangle$, as well as the finite-size effects of the fractal structure near $\tau_\mathrm{d}$ that affect $P(0,t)$, the AO relation holds in the reptation picture.

\begin{figure}[t]
		\centering 
		\includegraphics[width=\linewidth]{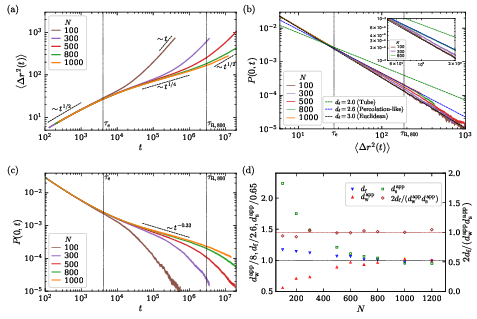} 
		\caption{\label{fig3} (a) $\langle \Delta r^{2}(t)\rangle$ as a function of $t$; dashed line indicates the power laws predicted by Eq.~\eqref{001} with $\nu =1/2$; (b) $P(0,t)$ versus $\langle \Delta r^{2}(t)\rangle$, where the dashed lines correspond to the Gaussian prediction (Eq.~\eqref{005}, $d_\mathrm{f}=3$), the percolation-like fractal structure ($d_\mathrm{f} = 2.6$) and the reptation expectation ($d_\mathrm{f}=2$), respectively, and intersect at the same point ($\langle \Delta r^{2}(t)\rangle=20.69\sigma^{2}$). Inset: zoomed-in view of selected regions. (c) $P(0,t)$ as a function of $t$; dashed line marks the power-law regime corresponding to MSD scaling $\langle \Delta r^{2}(t)\rangle \sim t^{1/4}$ for $\tau_\mathrm{e} <t < \tau_{\mathrm{R},800}$. (d) Fractal dimension $d_\mathrm{f}$, apparent spectral dimension $d_\mathrm{s}^{\mathrm{app}}$, apparent walk dimension $d_\mathrm{w}^{\mathrm{app}}$ and their ratio $2d_\mathrm{f}/(d_\mathrm{w}^{\mathrm{app}}d_\mathrm{s}^{\mathrm{app}})$ as functions of $N$, evaluated using the instantaneous logarithmic slopes at $t=10^4\tau$ for $N\geq 500$ and $t=10^5\tau$ for $N< 500$, in the time regime $\tau_\mathrm{e} <t \ll \tau_{\mathrm{R},800}$.}
	\end{figure}

	We now turn to entangled polymer melts, simulated using the LAMMPS package~\cite{plimpton1995fast} with the classical Kremer–Grest chain model~\cite{kremer1990dynamics}. The chain length ranges from $N=100$ to $N=1200$ with the number of chains varying between $200$ and $400$, and the monomer density is set to $\rho = 0.85\sigma^{-3}$, covering systems from unentangled to well-entangled melts~\cite{wang2012segmental,SM, behbahani2024relaxation}. It has been reported that the unique MSD power-law regime $\langle \Delta r^{2}(t)\rangle \sim t^{1/4}$, for entangled melts only becomes clearly visible for sufficiently long chains (approximately $N \gtrsim 600$)~\cite{wang2012segmental,behbahani2024relaxation}, as also shown in Fig.~\ref{fig3}(a). Note that, because the terminal relaxation time follows $\tau_\mathrm{d} \sim N^{-3.3}$, observing a clear $t> \tau_\mathrm{d}$ regime in the MSD is challenging for long chains, even with MD steps up to $10^{10}$ --- the current limit of computing power~\cite{behbahani2024relaxation}. This limitation, however, does not affect the accuracy of the data in the $t\ll \tau_\mathrm{d}$ regime, which is our primary interest. From the intersections of the MSD scaling regimes for $N=800$, we estimate $\tau_\mathrm{e} \approx 4.1\times10^3\tau$, which is essentially independent of chain length, $\tau_{\mathrm{R},800}\approx  3\times10^6\tau$, which increases with chain length (here $\tau$ is the MD time unit). 
	
	Figure~\ref{fig3}(b) shows $P(0,t)$ as a function of $\langle \Delta r^{2}(t)\rangle$. For $N=100$, the simulation data collapse onto the theoretical curve described by Eq.~\eqref{005}. As the chain length increases, deviations from this Gaussian behavior for $t >\tau_\mathrm{e}$ indicate the formation of a fractal structure by the surrounding chains. For $N<600$, we observe the complete process of sensing, moving within, and escaping from a finite-size structure. For longer chains, observing the full escape process becomes increasingly difficult. By examining the curvature of $P(0,t)$ versus $\langle \Delta r^{2}(t)\rangle$ and fitting the scaling, we find that for $\tau_\mathrm{e} < t\ll \tau_\mathrm{d}$, the behavior converges to $P(0,t) \sim \langle \Delta r^{2}(t)\rangle^{-1.3}$ with increasing chain length, indicating $d_\mathrm{f} \approx 2.6$ --- rather than the reptation prediction $d_\mathrm{f}=2$, as shown in Fig.~\ref{fig3}(c). Notably, for $N=800$ this power law persists clearly for $t$ appreciably larger than $\tau_{\mathrm{R},800}$. 
	
	In the same time window where $\langle \Delta r^{2}(t)\rangle \sim t^{1/4}$ for $\tau_\mathrm{e} <t < \tau_{\mathrm{R},800}$, we observe $P(0,t)\sim t^{0.33}$ over the same time window (see Fig.~\ref{fig3}(c)). Consistent with the previous two systems, we obtain $d_\mathrm{f}\approx 2.6$, $d_\mathrm{s}\approx 1.3$ and $d_\mathrm{w}\approx 4$, indicating that the AO relation also holds in well-entangled melts. These exponents suggest a percolation-like fractal structure rather than a tube. To clarify the chain-length dependence of these exponents, we evaluate the instantaneous logarithmic slopes at $t=10^4\tau$ for $N< 500$ and $t=10^5\tau$ for $N\geq 500$, in the time regime $\tau_\mathrm{e} <t \ll \tau_{\mathrm{R},800}$, as shown in Fig.~\ref{fig3}(d), based on the curvature of $P(0,t)$ versus $ \langle \Delta r^{2}(t)\rangle$. At these two times, Rouse dynamics already dominates even for the shortest chain $N=100$, so that $d_\mathrm{s} \approx 2 d_\mathrm{s}^{\mathrm{app}}$ and $d_\mathrm{w} \approx d_\mathrm{w}^{\mathrm{app}}/2$. Our results show that $d_\mathrm{f}$, $d_\mathrm{s}$ and $d_\mathrm{w}$ vary synchronously with chain length, crossing over from Euclidean values $3$, $3$ and $2$ for unentangled melts to $2.6$, $1.3$ and $4$ at $N\approx 600$ for well-entangled melts, while the AO relation remains valid throughout this crossover.

	Here we revisit the physical interpretation of Eq.~\eqref{002}, $S_t  \sim \langle \Delta r^{2}(t)\rangle^{d_\mathrm{f}/2}$, to gain deeper insight into entanglements. In a well-entangled melt, the chain can be coarse-grained into a Rouse chain of $N/N_\mathrm{e}$ blobs, each of size $N_\mathrm{e}b^{2}$, where $N_\mathrm{e}$ is the entanglement length and $b$ the Kuhn length. For times $t>\tau_\mathrm{e}$, a monomer and other monomers within the same blob effectively occupy the same spatial point from the perspective of visited-site counting; thus, $S_t$ can be interpreted as the number of distinct entanglements experienced by the tagged monomer up to time $t$. When the chain moves a distance comparable to its own size in Euclidean space, we have
	\begin{equation}
		\langle \Delta r^{2}(\tau_\mathrm{d})\rangle \approx N b^{2}
	\end{equation} 
	Within the reptation picture ($d_\mathrm{f} = 2$), the total number of entanglements encountered over such a displacement scales as
	\begin{equation}
	S_{\tau_\mathrm{d}}^{(\mathrm{rep})}  \sim N
	\end{equation} 	
	By contrast, using our measured $d_\mathrm{f}\approx 2.6$, we obtain
	\begin{equation}
	S_{\tau_\mathrm{d}}  \sim N^{1.3}
	\end{equation} 		
	which indicates a stronger chain-length dependence of entanglements than predicted by reptation.	
	
	We have introduced a model-independent method to directly quantify the fractal geometry experienced by a tracer particle and applied it to test the foundational premise of reptation theory. In obstructed systems and the slip-link model, our geometric probe confirms the existence of a fractal tube with $d_\mathrm{f}=1/\nu$, validating the core picture underlying reptation. Our approach captures the full process by which a tagged monomer enters, moves within, and eventually escapes from the tube, and we further confirm that including constraint release does not alter this static-obstacle picture. In entangled polymer melts, however, the same analysis reveals a looser fractal structure with $d_\mathrm{f}\approx 2.6$. Analysis of the spectral and walk dimensions $d_\mathrm{s}$ and $d_\mathrm{w}$ confirms the validity of the AO relation once the intrinsic multi-scale nature of entangled dynamics and the finite size of the fractal structure are properly taken into account. In contrast to the reptation picture for melts --- where a Rouse chain slides in a fractal structure with $(d_\mathrm{f}, d_\mathrm{s}, d_\mathrm{w}) = (2, 1, 4)$ --- our results point to a percolation-like scenario $(d_\mathrm{f}, d_\mathrm{s}, d_\mathrm{w}) \approx (2.6, 1.3, 4)$. These findings demonstrate that dynamic entanglements in polymer melts cannot be reduced to static obstacles; rather, entangled chains relax within a fractal geometry that is percolation-like, not tubular. Unlike conventional static percolation, this geometry emerges dynamically from the cooperative motion of chains and is intrinsically finite in size, indicating that entanglements are transient, self-organized topological constraints that continuously rearrange as the chain moves.

	\begin{acknowledgments}
		This work was supported by the National Natural Science Foundation of China (Nos. 22273013, 22473032) and the Guangdong Basic and Applied Basic Research Foundation (No. 2024A1515010027). 
	\end{acknowledgments}
	
	\bibliography{apstemplate}
	
\end{document}